\documentclass{IEEEtran}
\usepackage{amsmath,amssymb}

\usepackage{subfigure}
\usepackage{graphicx}
\usepackage{float}
\usepackage{caption}
\usepackage{slashbox}
\usepackage{array}
\usepackage{bm}

\usepackage[ruled,vlined,linesnumbered]{algorithm2e}
\usepackage{setspace}
\usepackage[usenames]{color}
\usepackage{cases}

\usepackage{CJK}
\usepackage{tabularx}
\usepackage{multirow}
\usepackage{xcolor}
\usepackage{colortbl,booktabs}

\usepackage{amsmath,amssymb}
\usepackage{epstopdf}
\usepackage{subfigure}
\usepackage{graphicx}
\usepackage{float}
\usepackage{caption}
\usepackage{slashbox}
\usepackage{longtable}
\usepackage{rotating}
\usepackage{multirow}
\usepackage{array}
\usepackage{bm}
\usepackage{algorithmic}

\usepackage[ruled,vlined,linesnumbered]{algorithm2e}
\usepackage[usenames]{color}
\usepackage{amscd}
\usepackage{multirow}
\usepackage{bigdelim}
\usepackage{caption}

\definecolor{mygray}{gray}{.9}
\definecolor{myblue}{RGB}{135,206,250}
\definecolor{mybluegray}{RGB}{119,136,153}

\begin{document}

\markboth{IEEE Network, Vol. XX, No. YY, Month 2019}
{Cao, Li, Zhang, \& $et$ $al$: Blockchain \ldots}

\title{When Internet of Things Meets Blockchain: Challenges in Distributed Consensus}

\author{Bin Cao, Yixin Li, Lei Zhang, Long Zhang, Shahid Mumtaz, Zhenyu Zhou and Mugen Peng

\thanks{Bin Cao (email: {\tt caobin@cqupt.edu.cn}), Yixin Li and Long Zhang are with the Chongqing University of Posts and Telecommunications of China, College of Communications and Information Engineering and Chongqing Key Lab of Mobile Communications Technology, Bin Cao is also with the State Key Laboratory of Integrated Services Networks (Xidian University) and the Beijing University of Posts and Telecommunications of China, Institute of Network Technology. Lei Zhang is with the School of Engineering, University of Glasgow, Glasgow, G12 8QQ, U.K. Shahid Mumtaz is with the Institute of Telecommunications, Portugal. Zhenyu Zhou is with the University of North China Electric Power, College of Electrical Engineering. Mugen Peng is with the Beijing University of Posts and Telecommunications of China, College of Communications and Information Engineering.}
\thanks{This work was supported in part by the National Natural Science Foundation of China (61701059, 61831002), the State Major Science and Technology Special Projects (2017ZX03001025-06), the Beijing Natural Science Foundation under Grant No. JQ18016, and the Eighteenth Open Foundation of State Key Lab of Integrated Services Networks of Xidian University (ISN20-05).}}

\maketitle


%

\begin{abstract}

Blockchain has been regarded as a promising technology for Internet of Things (IoT), since it provides significant solutions for decentralized network which can address trust and security concerns, high maintenance cost problem, etc. The decentralization provided by blockchain can be largely attributed to the use of consensus mechanism, which enables peer-to-peer trading in a distributed manner without the involvement of any third party. This article starts from introducing the basic concept of blockchain and illustrating why consensus mechanism plays an indispensable role in a blockchain enabled IoT system. Then, we discuss the main ideas of two famous consensus mechanisms including Proof of Work (PoW) and Proof of Stake (PoS), and list their limitations in IoT. Next, two mainstream Direct Acyclic Graph (DAG) based consensus mechanisms, i.e., the Tangle and Hashgraph, are reviewed to show why DAG consensus is more suitable for IoT system than PoW and PoS. Potential issues and challenges of DAG based consensus mechanism to be addressed in the future are discussed in the last.

\end{abstract}

\begin{IEEEkeywords}
Consensus Mechanism, Blockchain, Internet of Things, Direct Acyclic Graph, Tangle.
\end{IEEEkeywords}

\IEEEpeerreviewmaketitle

\section{Introduction}

Internet of Things (IoT) has been identified as one of the most disruptive technologies of this century. It has attracted much attention of society, industry and academia as a promising technology that can enhance day to day activities, the creation of new business models, products and services, and as a broad source of research topics and ideas. Although the first idea of IoT emerged no more than two decades ago and many IoT ecosystems have been generated since then, some unsolved and important issues are still remained as follows:\begin{itemize}
 \item Trust: IoT cloud servers are closed systems. For one thing, the service providers have the ability to illegally control IoT devices. For another, it is hard to build the cooperation and trust relationship among different IoT business agencies;
 \item Security: the IoT data center is vulnerable since it is easy to be attacked by hackers using Distributed Denial of Service attack (DDoS), and when it happens, all IoT service may be affected due to the centralized topology;
 \item Overhead: current centralized model has a high maintenance cost, i.e., it is costly to timely update the softwares of millions of IoT devices;
 \item Scalability: the poor scalability of the centralized topology cannot meet the needs of massive IoT devices connection, i.e., a large delay might be caused by a surge of service requests.

\end{itemize}

As a brand of new distributed ledger technology (DLT), blockchain is originally designed for digital currency Bitcoin in 2009 \cite{2-bitcoin}. With decades of operation in a decentralized network, Bitcoin did not encounter serious security incidents. This can be largely attributed to the advantage of consensus mechanism, which uses the computing power of whole network to ensure the immutability of the data. As such a security decentralization solution, blockchain is expected to transform IoT ecosystems by making them smart and more efficient. According to IDC (International Data Corporation) report, by 2019, 20\% of IoT deployments will have basic levels of blockchain enabled services \cite{6-IDC}.

\subsection{What is Blockchain}

Blockchain is a peer-to-peer (P2P) distributed ledger technology for establishing trust and consensus in decentralized networks. On the one hand, to address the challenges in trustless distributed environment\footnote{Refer to the Byzantine Generals Problem \cite{5-consensus}.}, consensus mechanism is adopted in blockchain in a decentralized way to reach the agreement for transactions among individual users. On the other hand, using digital signature and hash algorithm based encryption, security can be assured in the decentralization blockchain system \cite{4-blockchain}.

Blockchain ledger has three basic concepts: transaction, block and chain. The ``transaction" in blockchain is not restricted for trading, in fact, all the valuable information can act as a transaction to be broadcast in blockchain network. The blocks are storage units to record transactions, which are created and broadcast by those users authorized by consensus mechanism. Each block is identified uniquely by its hash value, which is referenced by the block came after it. This establishes a link between the blocks, thus creating a chain of blocks namely ledger. With the blocks accumulate sequentially in consensus process, the cost of attack and malicious modification would be increased exponentially \cite{2-bitcoin}.

 \begin{figure*}[htbp]
\begin{center}
 \includegraphics[width=17cm]{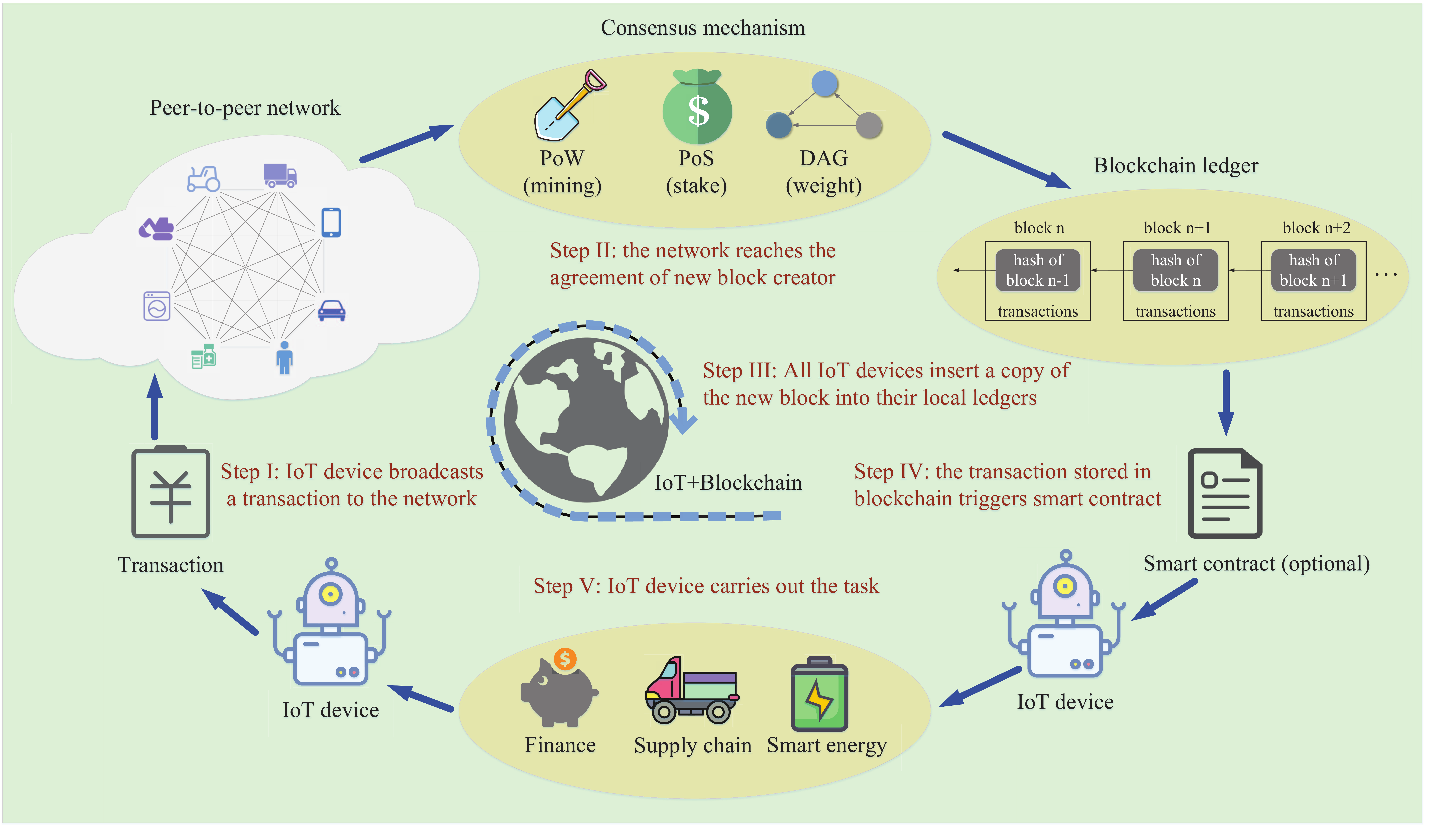}
 \end{center}
 \caption{An example of implementing blockchain in IoT system}
\label{Fig5}
\end{figure*}

\subsection{Advantages of Blockchain for IoT}

Firstly, using blockchain based decentralization, the burden of hot spot and the probability of Single Point of Failure (SPF) can be reduced significantly. Secondly, consensus mechanism and encryption algorithm in blockchain can be leveraged to strengthen IoT security. In addition, by using smart contact \cite{2-smart}, IoT devices can carry out trading and execute actions autonomously. Besides, as a public distributed ledger where stored information can be audited by all the users, blockchain provides a trust platform for IoT business cooperation.

\subsection{IoT and Blockchain Integration}

Currently, the implementation of IoT and blockchain is on the agenda in industry and there are already promising solutions and initiatives in several areas. In supply chain industry, \cite{2-smart} provides a blockchain enabled supply chain model. In this model, the infomation stored in the blockchain can serve as a log of delivery for container shipments. All the movement of container from source to destination can be tracked by any supply chain entities, so that the shipment delay can be minimized and the missing asset can be tracked accurately. In healthcare domain, \cite{4-Application1} provides a user-centric model for processing personal health data using blockchain network, ensuring the data ownership of individuals, as well as data integrity. By enforcing access control policies, the system makes sure that users can handle their personal data without worrying about the privacy issues. Besides, blockchain is also available in the other IoT applications, such as remote software updates and insurance for vehicle \cite{6-Application2}.

Particularly, blockchain plays an important role in energy trading for IoT applications in energy Internet. Nowadays, there exist some blockchain technologies which have investigated how to promote energy sharing among IoT devices to increase efficiency of energy utilization. Taking the Internet
of Vehicles (IoVs) as an example, the electric vehicles have the ability to absorb excessive energy during the non-peak area and provide energy as distributed generators during the peak period. To enable secure energy trading, \cite{enabling} proposes a localized P2P electricity trading framework, in which consortium blockchain is exploited to improve the security of transaction without relying on a third party. To improve the trading efficiency, \cite{consortium} proposes a credit-based payment scheme, which supports the fast and frequent trading among energy nodes by establishing virtual credit banks. Besides, some digital currency has been presented for renewable trading based on blockchain, such as ``Specoin" \cite{secure}.

As shown in Fig. 1, to operate a blockchain enabled IoT system, the main steps are illustrated as follows: (i) All IoT devices operate on the same blockchain network; (ii) A IoT device generates a transaction for payment (or recording significant information), and broadcasts it to the network; (iii) The IoT devices receive the information and transactions in the network and validate them; (iv) All IoT devices perform hash algorithm to elect a winner whose candidate block will be broadcast and validated as a new block. (v) All IoT devices insert the identical copy of the new block into their local ledgers. (vi) The transaction stored in blockchain ledger triggers the smart contract\footnote{Smart contract is only an option in this circle, which is an application on top of blockcahin, the IoT devices may use blockchain for many other applications without relying on smart contract.} in IoT device. (vii)IoT device carries out a specific task, i.e., the movement of container in supply chain scenario, power supplying in smart energy scenario.

According to Fig. 1, we can see that consensus mechanism is the cornerstone in blochchain enabled IoT system, which builds a bridge between the raw data from infrastructure and the confirmed information for performing various applications. Therefore, the goal of this work is to clarify the challenges of consensus mechanism for blockchain enabled IoT systems. We illustrate the main idea of different types of consensus mechanisms and list their advantages and disadvantages in IoT ecosystem, then discuss some possible research directions of Direct Acyclic Graph (DAG) based consensus mechanisms.

The rest of this article is organized as follows. In Section II, we introduce the main idea of consensus mechanism, including Proof of Work (PoW), Proof of Stake (PoS) and DAG, and consider their practicability for IoT. In Section III, we review two existing DAG based consensus (Tangle and Hashgraph) and demonstrate their advantages in IoT through performance comparisons. In Section IV, we discuss some research directions of DAG based consensus. Conclusions are drawn in Section V.

\section{Consensus Mechanism in Blockchain}

In this section, we discuss different types of consensus mechanisms in blockchain, and consider whether the design criteria of corresponding consensus mechanism can meet the needs of IoT.

Consensus mechanism plays an indispensable role in blockchain to resolve the trust concern by answering the question ``who will be the one has the right to insert the next block into blockchain". With consensus mechanism, the information can be announced orderly to all users without involvement of the third party. Nowadays, various consensus mechanisms have been proposed, PoW and PoS are the most widely used ones. However, the two consensus mechanisms based traditional blockchains face significant challenges when apply to IoT system. We introduce DAG based consensus mechanism as an effective solution.

\subsection{Blockchain $1.0:$ Proof of Work}

PoW is proposed in the original blockchain application (e.g., Bitcoin). The core idea of PoW is the competition of computing power \cite{2-bitcoin}, the node performing the consensus mechanism (called miner) uses its computing resource for hashing operation to compete for the right to generate the new block with bonuses. The winner is the first one who obtains a hash value lower than the announced target. On the one hand, the computing difficulty in PoW must be high enough for preventing forking \cite{4-blockchain}. But on the other hand, the high computing difficulty would cause the deteriorated and meaningless energy consumption. Noted that the available resource of IoT devices is very limited. Therefore, PoW is not a good option for IoT system.

\subsection{Blockchain $2.0:$ Proof of Stake}

Unlike PoW that relies on computing capability, coin age is used in PoS blockchain to avoid the high computational complexity of hash operation (e.g., Nxt\cite{11-nextcoin}). The coin age of an unspent transaction output\footnote{The output of a transaction includes destination address and the amount of coin.} is equal to its value multiplied by the time period after it was created. In PoS, a higher coin age will lead to a higher probability for the node to win the right of creating a new block, and in turn the coin age would be consumed (reset as zero) when the owner wins. Since winning probability is directly determined by coin age, PoS is beneficial for the wealthy miner, and might cause oligopolies or near-monopolies, then result in the generation of powerful third party. From this sense, the PoS consensus mechanism may not fit well to establish a smart distributed IoT systems.

\subsection{Limitations of PoW and PoS for IoT}

PoW and PoS are two typical traditional consensus mechanisms that work on a ``single chain" (forking is illegal) architecture. To avoid forking and maintain a single version of blockchain ledger among all users, the consensus mechanism must slow down the access rate of new blocks. This might cause some significant bottlenecks in applying to IoT system. (i) \emph{Resource consumption}: to slow down the access rate of new blocks and prevent blockchain network from attack, the traditional consensus process will consume much resource (i.e., computing power in PoW, coinage in PoS), which is too costly for the resource-limited IoT devices. (ii) \emph{Transaction fee}: transaction fee is needed in traditional consensus mechanism to feed the miners, which might cause a heavy burden in the IoT system where most of tradings are micropayments. (iii) \emph{Throughput limitation}: since the capacity of a new block is limited, Transaction Per Second (TPS) is limited to dozens usually (e.g., 7 TPS in Bitcoin and 20 to 30 TPS in Ethereum), which is unable to respond to the exponential growth of IoT devices. (iv) \emph{Confirmation delay}: due to the low access rate of new blocks, the confirmation delay is too long for IoT applications (e.g., 60 minutes in Bitcoin and 3 minutes in Ethereum).

\subsection{BlockChain $3.0:$ Direct Acyclic Graph}

DAG architecture and its consensus mechanism is proposed to overcome the shortcomings of traditional consensus for IoT. Some typical DAG consensus processes are shown in Fig. 2 and Fig. 3. DAG based consensus mechanism allows users to insert their blocks into the blockchain at any time, as long as they process the earlier transactions. In this way, many branches would be generated simultaneously, which is called as forking. This phenomenon is usually regarded as an issue in many traditional consensus process since it would cause ``double-spending" \cite{2-bitcoin}. However, DAG based consensus mechanism design innovative protocol and algorithm (detailed in next section) to address the double-spending problem, and allow any new arrival transactions access the blockchain network in a forking topology. As a result, the confirmation rate and TPS will not be limited anymore. Moreover, since the data stored in DAG is protected by massive forking blocks, the resource consumption can be very low for a single user to create a new block. Accordingly, professional miner disappears and low or no transaction fee is possible, which is critically important to IoT ecosystem.

\section{Typical DAG based Consensus}

In this section, we introduce the consensus mechanism in Tangle and Hashgraph, respectively, which are the two typical DAG based consensus.

\subsection{The Tangle}

 \begin{figure}[t]
\begin{center}
 \includegraphics[width=8cm]{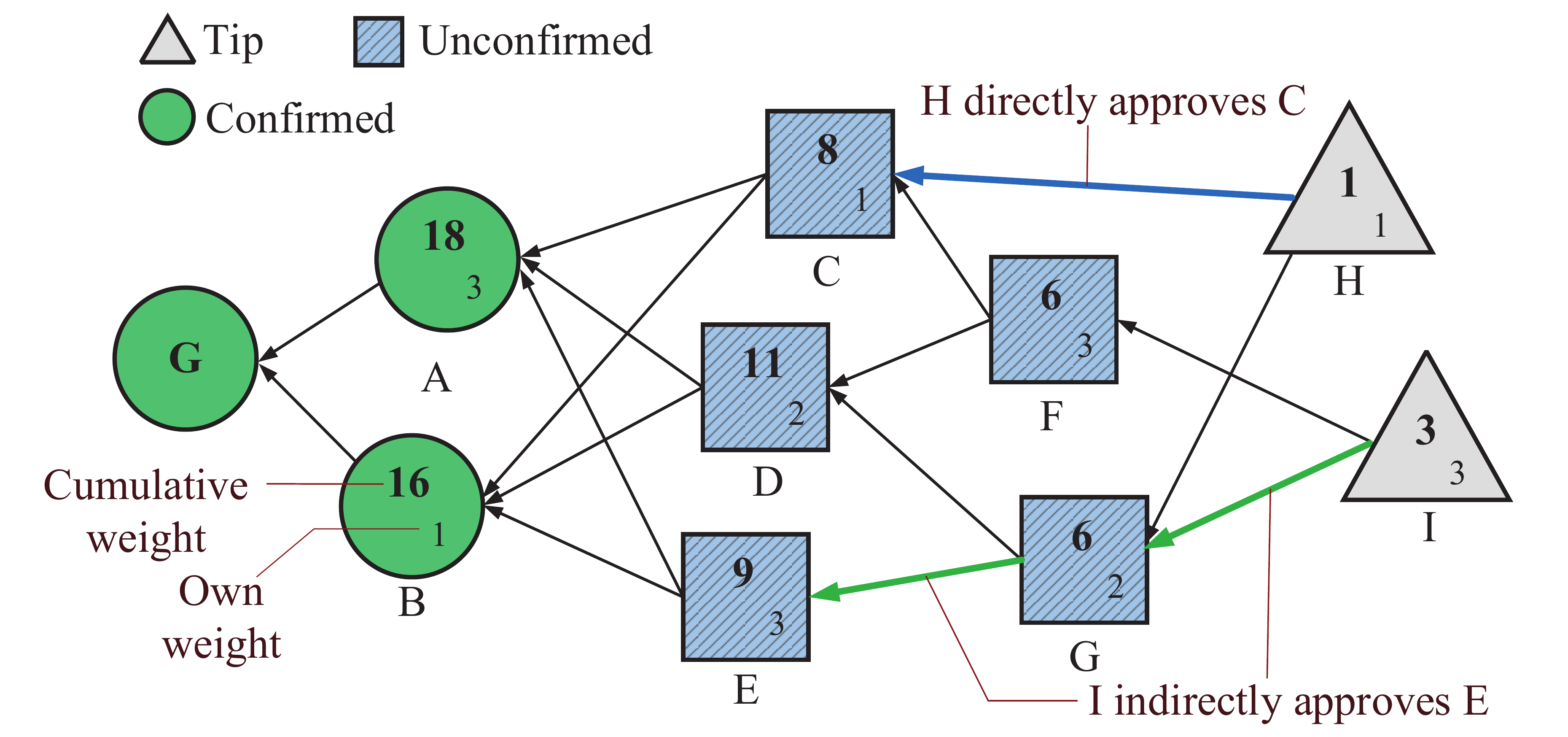}
 \end{center}
 \caption{An example of Tangle}
\label{Fig5}
\end{figure}

Tangle is the mathematical foundation of IOTA \cite{7-tangle}, a cryptocurrency for the IoT industry. As shown in Fig. 2, Tangle is a DAG based distributed ledger for recording transactions. It allows different branches to eventually merge into the chain, resulting in a much faster overall throughput. In Tangle, to access the ledger as a new vertex for storing a transaction, it has to approve a number of tips (typically two \cite{7-tangle}). Thanks to this, the higher arrival rate of new transactions, the faster earlier transactions can be confirmed. On the other hand, since tips are the childless vertexes in Tangle, the new vertex selects tips and covers them could limit the branch to a reasonable scale. Moreover, since the workload to create a new vertex is light, all users can issue their transactions at any time without transaction fee, which is critical to the IoT application scenarios.

The consensus in the Tangle relates to cumulative weight. As shown in Fig. 2, the cumulative weight of a specific transaction is the sum of a vertex's own weight (proportional to the PoW that the issuing node invested into it \cite{7-tangle}) and the overall weights of the vertices directly and indirectly approve it. Since the transactions stored in Tangle are secured by computing power, the cumulative weight of a transaction means its validity in the network and act as a decisive criteria to address double-spending problem.

In order to issue a new transaction and let the other users in the whole system accept it (i.e., win enough cumulative weight to reach an agreement for the consensus), the main procedures are listed as follows. (i) A user creates a unit as a candidate vertex in the DAG graph to store its transaction and uses its private key to sign the transaction. (ii) The user selects two tips with no-conflict according to a Markov Chain Monte Carlo (MCMC) algorithm \cite{7-tangle}, and adds the hash of the selected tips into its storage unit. (iii) The user finds a nonce to solve a cryptographic puzzle to meet the difficulty target. It is similar to PoW but with a very low difficulty-of-work, which can avoid spamming. After that, the user broadcasts the storage unit to others. (iv) When the other users receive it, they should check whether it is legal or not based on the digital signature and PoW based nonce. Successfully checked new storage unit would be added as a new tip in the Tangle, and waits for confirmation through direct approval and indirect approval till its cumulative weight reaches the predefined threshold.

In a public ledger, building forking (or branch) and redoing the work is the only way to tamper with data and conduct double-spending. To address this problem, the single chain based consensus mechanism (e.g., PoW) use the longest chain as the criterion. To this end, to guarantee and maximize the own profit, a rational user should choose the longest chain to work when forking occurs. The reason is that the longest chain has the lowest probability to be orphaned. Similarly, the Tangle uses the MCMC tip selection algorithm to select the branch with the largest overall cumulative weight. Moreover, with the assistance of distributed and parallel approval in Tangle, the overall computing capability of honest users in large scale IoT system could be powerful to prevent double-spending, where the branch generated by an attacker is hard to outweigh the honest one. Meanwhile, any single user does not need to consume much power on computing for security.

\subsection{Hashgraph}

 \begin{figure}[t]
\begin{center}
 \includegraphics[width=9cm]{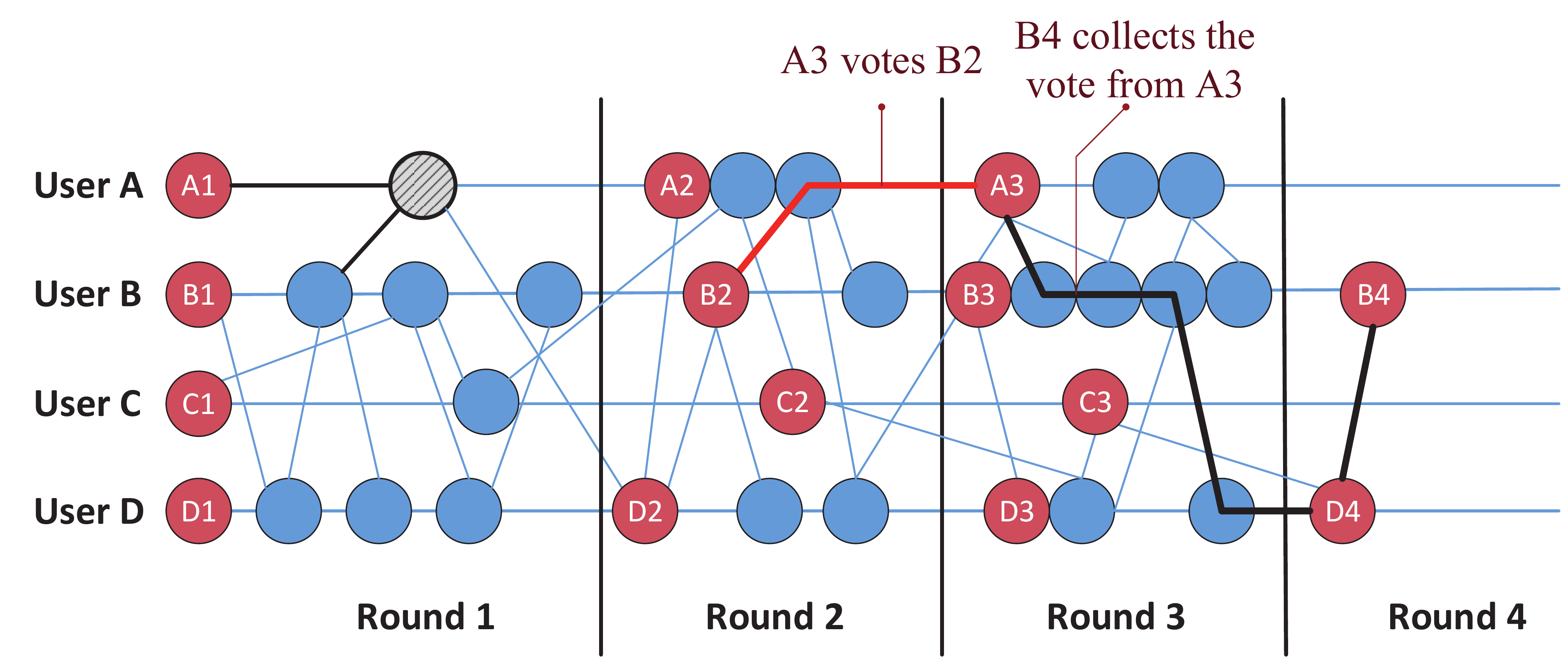}
 \end{center}
 \caption{An example of Hashgraph}
\label{Fig5}
\end{figure}

Hashgraph \cite{9-hashgraph} is proposed for replicated state machines with guarantee of Byzantine fault tolerance, it is asynchrony, decentralization, no PoW, eventual consensus with probability one, and high speed in the consensus process. Gossip protocol and virtual voting are two key elements in Hashgraph. Using gossip protocol, every transaction will be known by all users. After that, the agreement of the order of transactions will be reached through virtual voting algorithm. In order to get a better understanding of Hashgraph, we will briefly introduce how gossip protocol and virtual voting work.

\newcommand{\tabincell}[2]{\begin{tabular}{@{}#1@{}}#2\end{tabular}}
\renewcommand\arraystretch{2}
\begin{table*}[!t]
\centering
\scriptsize
\caption{Comparisons of PoW, PoS and DAG based Consensus}
\begin{tabular}{|
c|%
l|%
l|%
l|%
l|
l|}

\arrayrulecolor{mybluegray}\hline

  & Bitcoin \cite{2-bitcoin} & Nxt \cite{11-nextcoin} & Tangle \cite{7-tangle}  & Hashgraph \cite{9-hashgraph} \\
\arrayrulecolor{mybluegray}\hline
\tabincell{l}{Byzantine fault tolerance} & \tabincell{l}{ $<$51\% of all computing \\ resource }
                                    & \tabincell{l}{ $<$ 1/3 of total assets}
                                    &  \tabincell{l}{$<$51\% of all computing resource \\using MCMC tips selection}
                                    & \tabincell{l}{Dishonest participants $<$ 1/3} \\

\arrayrulecolor{mybluegray}\hline
Transaction fee & \tabincell{l}{ 0.0001 BTC }
            & \tabincell{l}{ 1 Nxt}
            & \tabincell{l}{Zero}
            & \tabincell{l}{Zero}\\

\arrayrulecolor{mybluegray}\hline
\tabincell{l}{Resource requirements}   & \tabincell{l}{Enormous computing power}
            & \tabincell{l}{Coin age}
            & \tabincell{l}{ Low computing power}
            & \tabincell{l}{Low computing power and bandwidth}\\

\arrayrulecolor{mybluegray}\hline
Throughput  & \tabincell{l}{7 TPS }
            & \tabincell{l}{4 TPS }
            & \tabincell{l}{No technical up bound}
            & \tabincell{l}{ $2.5 \times 10^5$ TPS}\\
\arrayrulecolor{mybluegray}\hline
\tabincell{l}{Confirmation delay}   & \tabincell{l}{  60 mins }
            & \tabincell{l}{ 10 mins  }
            & \tabincell{l}{Depend on transaction arrival rate}
            & \tabincell{l} {Subject to communication frequency}\\
\arrayrulecolor{mybluegray}\hline
Finality    & \tabincell{l}{Six cumulative blocks \\ at least }
            & \tabincell{l}{Ten cumulative blocks \\ at least }
            & \tabincell{l}{Cumulative weight reaches \\ confirmation threshold }
            & \tabincell{l}{Seen by all the famous witnesses \\ in a latter round}\\
\arrayrulecolor{mybluegray}\hline
\tabincell{l}{Unique features}  &  \tabincell{l}{$\bullet$ Competition for mining\\
                                  $\bullet$  PoW}
                & \tabincell{l}{$\bullet$ The miner of the next\\ block
                 are predictable\\
                               $\bullet$ PoS }
                &  \tabincell{l}{$\bullet$ Offline transactions \\
                               $\bullet$ Quantum Immune\\
                               $\bullet$ DAG}
                &  \tabincell{l}{$\bullet$ Proof of Asynchronous Byzantine\\
                              fault tolerance \\
                               $\bullet$ Gossip to gossip and Virtual voting\\
                            $\bullet$ DAG \\}\\

\arrayrulecolor{mybluegray}\hline
\tabincell{l}{Major drawback} &  \tabincell{l}{High resource consumption \\ (hash complexity)}
                & \tabincell{l}{Centralization concern \\ (coin age)}
                &  \tabincell{l}{$\bullet$ The large confirmation delay\\
                        in low trading traffic load \\
                     $\bullet$ Centralization concern \\ (when coordinator involves)}
                 &  \tabincell{l}{The large confirmation delay caused\\
                    by low communication frequency\\
                    (gossip protocol) }\\

\arrayrulecolor{mybluegray}\hline

\end{tabular}
\end{table*}

According to gossip protocol, in a fixed interval, each user in Hashgraph should randomly choose another one to announce all the transactions it knows. For example, the shadow unit in Fig. 3 represents user B sends some information to user A that A does not know, so A creates the event which links A and B to store the unknown information. In this way, every event will be known by all participants eventually. Note that gossip protocol is a low-cost method, the overhead to exchange a storage unit is very small, which includes positional information (3 to 6 bytes), signature (64 bytes) and transactions within the unit (about 100 bytes).

To achieve the consensus, the system needs to select the ``famous witnesses" through virtual voting (all users perform the voting algorithm based on the graph connectivity). The famous witnesses are elected from witnesses which are the first events in each round (the red units in Fig. 3). An electing process includes voting and checking. As shown in Fig. 3, the witnesses in round 3 vote for the witnesses in round 2. Then, the witnesses in round 4 will collect the votes in round 3. If the voting in round 3 and checking in round 4 succeed, the witnesses in round 2 would become famous. The events in round 1 voted by the famous witnesses in round 2 will be confirmed. The creation time of the confirmed events will be accepted by all users, which acts as a proof to prevent double-spending.

\subsection{Comparisons}

To demonstrate the advantages and limitations of DAG based consensus for IoT, we compare its performance with two mainstream consensus mechanism in Table I.

These comparisons reflect that DAG based consensus mechanisms are more suitable for large-scale IoT than PoW and PoS. Specifically, DAG based consensus has the lower transaction fee, resource consumption, and it can achieve a much higher transaction throughput. However, some limitations are still remained in DAG based consensus mechanisms, e.g., centralization concern in Tangle. Moreover, the confirmation delay of DAG consensus would be affected by traffic load significantly, especially when the traffic load in practical IoT scenario changes over time. Hence, to apply DAG based consensus, the mentioned issues but not limited on these should be addressed.

\section{Challenges of DAG based Consensus}

Although DAG based consensus mechanism has many advantages, as an emerging technology, it is still far from perfect to be widely used in IoT systems. Some main issues of DAG based consensus are open to be explored.

\begin{figure}[t]
\begin{center}
 \includegraphics[width=9cm]{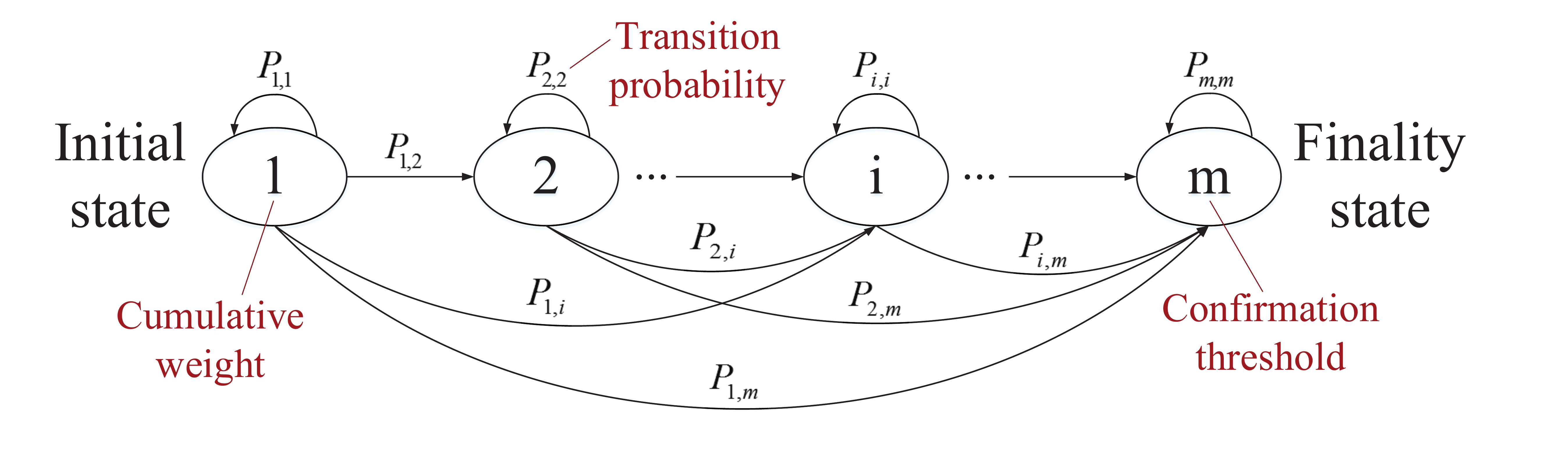}
 \end{center}
 \caption{Markov chain model for the consensus process of a new transaction}
\label{Fig6}
\end{figure}

\subsection{Analysis Model}

Design a generalized theoretical mathematical model is important to analyze the performance of DAG based consensus machanism. In \cite{7-tangle}, the authors analyze the speed of the cumulative weight typically grow in the stationary high load regime, it provides some qualitative and quantitative insights into the consensus process of the Tangle. In \cite{14-Equilibria}, the authors prove the existence of (``almost symmetric") Nash equilibria in a DAG-valued stochastic process where a part of players try to optimize their strategies.

Considering the features of consensus process, we believe that an analytical model using Markov chain is a promising approach. The formulation of an Markov chain model for the consensus process of a new transaction is shown in Fig. 4. The model uses the cumulative weight introduced in Tangle as the confirmation criterion. Accordingly, we can analyze the $N$-step transition probability from the current state to the finality state. As a result, the increasing rate of cumulative weight, TPS and confirmation delay can be analyzed in a theoretical approach.

One of the most significant and remaining problem of the Markov Chain based model is how to capture the transition probabilities matrix, especially in the large network scale with the huge number of system states. Moreover, the transition probability is also strongly affected by the design criteria of consensus mechanism, e.g., they are totally different between Tangle and Hashgraph. Therefore, the Markov Chain based model needs to be optimized in the future work.

\subsection{The Low Bound Limitation}

As we mentioned before, there is no technical up bound of throughput in DAG based consensus process. However, in practical IoT scenario, it is impossible that the new transaction arrives quickly and steadily all the time. Taking bicycle-sharing application as an example, there are very few transactions at night. In this case, the confirmation delay could be quite large.

In order to show the impact of arrival rate (defined as $\lambda$) on the consensus process, we conduct a simple simulation based on the Markov chain model in section A. In Fig. 5, we can see clearly that when the arrival rate of new transaction is low, the cumulative weight would increase slowly. Since the confirmation of a transaction is determined by its cumulative weight \cite{7-tangle}, as a result, the confirmation delay would be very large when the arrival rate is low.

 \begin{figure}[t]
\begin{center}
 \includegraphics[width=8cm]{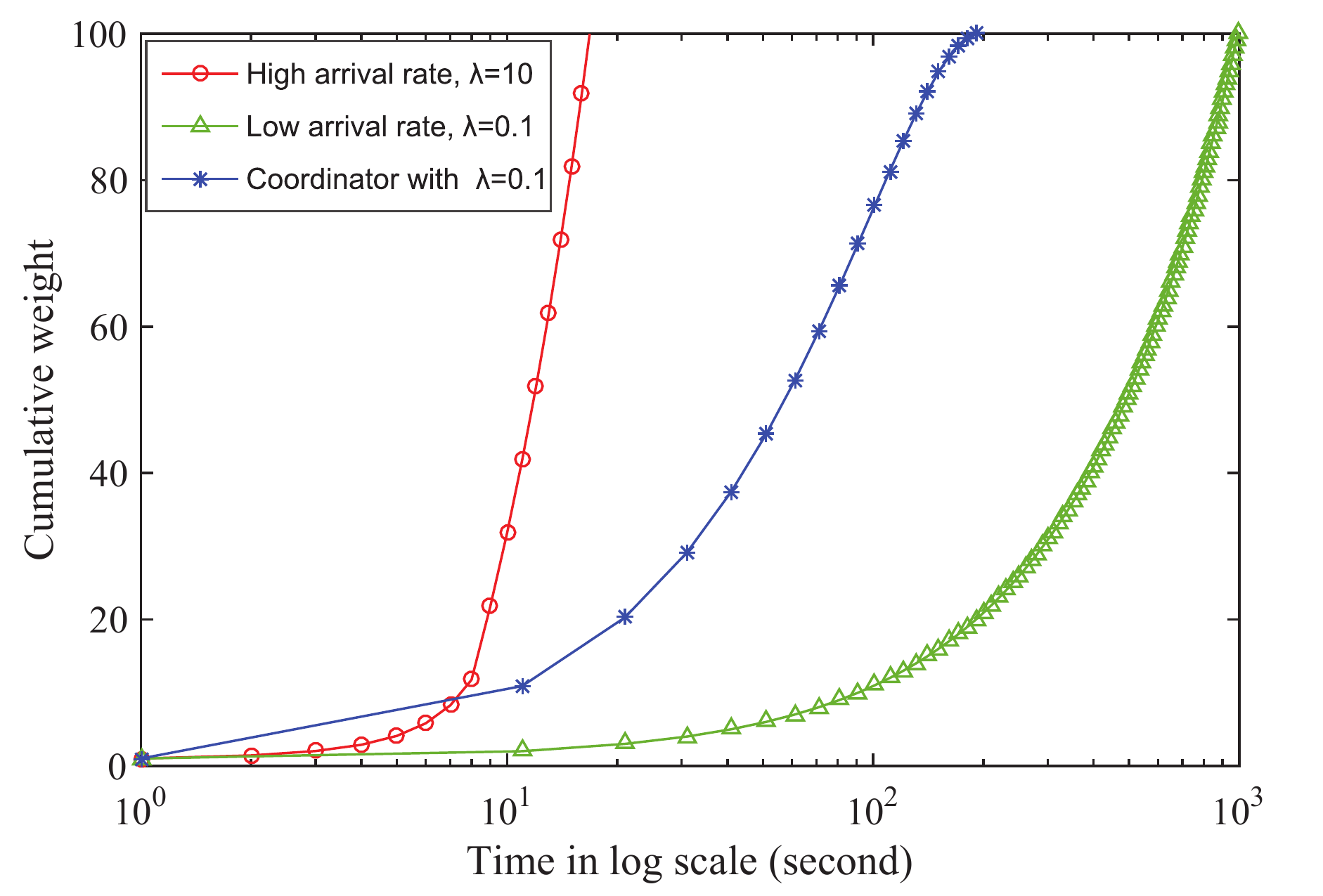}
 \end{center}
 \caption{Cumulative weight growth curve in different regimes}
\label{Fig5}
\end{figure}

To this end, coordinator is involved in DAG based consensus process to improve confirmation rate in low trading traffic load regime. The coordinator is an entity controlled by a third party, which issues zero-value transactions to process unconfirmed transactions. In Fig. 5, we can see that with the assistance of coordinator, the cumulative weight increases more quickly in the low arrival rate situation. On the one hand, this solution could resolve the large confirmation delay issues in the low arrival rate situation. On the other hand, centralization problem might be incurred, since the coordinator is a third party, which disobeys the basic rule of blockchain. Due to this, coordinator can only be used in private or closed situations, i.e., consortium blockchain.

\subsection{Mobile Blockchain}

It is nature to assume that typical IoT devices are wireless connected. In many researches on consensus process (i.e., Tangle and Hashgraph), communications are assumed wired or perfect. However, due to the wireless channel fluctuation, the communication might be a bottleneck for the blockchain enabled IoT systems. We discuss the challenges related to communication in blockchain enabled IoT systems from different layers.

\subsubsection{Lower layer}

In physical layer, some fundamental metrics such as signal-to-interference-plus-noise ratio (SINR) and communication throughput should be analyzed to show how the wireless communication quality affect/constrain the blockchain-enabled IoT system deployment (e.g., node distribution), protocols (e.g., size of block, frequency of transactions) and confirmation delay, etc. On the other hand, given a transaction throughout bound in blockchain (e.g., one block in every 10 minutes as defined in Bitcoin \cite{2-bitcoin}), it is valuable to know how to deploy the IoT devices that can optimally meet this bound. Another challenge comes from the fact that IoT devices might be massively connection, which has been identified as one of the main features for fifth-generation (5G) wireless communication. The trade-off between the system overhead and security performance will be an interesting topic to be explored. In addition, physical links and access control protocol will influence the communication performance in terms of throughput and latency, which might be two factors that may pose extra bottleneck to the consensus process. Finally, joint wireless and consensus mechanism design to maximize the overall security level is of interest from the system level.

\subsubsection{Upper layer}

In route layer, considering the memory space and processing capacity of IoT devices are normally constrained, the deteriorated delay in bottleneck would affect the consensus process (i.e., the congested IoT device might be regarded as the a ``lazy" node erroneously \cite{7-tangle}). Therefore, an efficient routing protocol in blockchain enabled IoT system should prefer the resourceful IoT devices to propagate transactions. Meanwhile, in Transmission Control Protocol (TCP) layer, a protocol should be designed to meet the specific QoS needs of blockchain network. Especially, when a transaction failure occurs, the protocol should identify the exact reason. If the transaction failure is caused by transmission error or timeout rather than consensus mechanism, the retransmission should be performed by the protocol for correction and recovery.

\subsection{Blockchain Strategy Optimization}

In DAG based consensus mechanism, every participant is also a transaction verifier to store and update the ledger in a distributed manner. Since most IoT devices are limited for power and memory, the energy saving and caching strategy should be well designed to lighten and balance the workload of each user. For instance, a resource optimization strategy, which allows resource-limited IoT devices to issue transactions only, resourceful IoT devices to process the transactions and generate blockchain, can be developed. Meanwhile, due to the selfishness and rationality, some incentive mechanisms should be performed to motivate the suitable IoT devices to participate into consensus process. In order to let IoT devices make the optimal strategy in a distributed manner, game theory is a nature selection. For example, in \cite{12-niyato}, the authors propose an auction based approach for PoW offloading in mobile blockchain.

\section{Conclusions}

In this article, we have introduced the concept of blockchain and the benefits of using it into the IoT systems. We start from illustrate the main ideas of consensus mechanism including PoW, PoS and DAG, and discuss their advantages and limitations for IoT. Two DAG based consensus mechanisms, i.e., Tangle and Hashgraph are introduced. We also compare the main characteristics of PoW, PoS, and DAG. Furthermore, we present a visible simulation results to show the impact of transaction arrival rate on consensus process in DAG based blockchain, and reveal its low bound limitation. Challenges for the DAG based consensus mechanism to use in the IoT system are summarized from analysis model, major drawback, mobile blockchain and optimization strategy.

\begin{IEEEbiographynophoto}{BIN CAO} (caobin65@163.com)
received his Ph.D. degree in communication and information systems from the University of Electronic Science and Technology of China, Chengdu, China in 2014. Currently, he is an associate professor with the College of Communication and Information Engineering, Chongqing University of Posts and Telecommunications, Chongqing, China. His research interests include blockchain, internet of things, cooperative communications, wireless network virtualization, and mobile cloud computing.
\end{IEEEbiographynophoto}

\begin{IEEEbiographynophoto}{YIXIN LI} (liyixinggg@163.com)
is pursuing his Master degree at the School of Communication and Information Engineering, Chongqing University of Posts and Telecommunications, Chongqing, China. His research interests include blockchain and internet of things.
\end{IEEEbiographynophoto}

\begin{IEEEbiographynophoto}{LEI ZHANG} (lei.zhang@glasgow.ac.uk)
received his Ph.D. from the University of Sheffield, U.K. He is now a Lecturer at the University of Glasgow, U.K. His research interests broadly lie in the Communications and Array Signal Processing, including radio access network slicing (RAN slicing), wireless blockchain systems, new air interface design, Internet of Things (IoT), multi-antenna signal processing, massive MIMO systems, etc. Dr Lei Zhang also holds a visiting position in 5GIC at the University of Surrey. He is an associate editor of IEEE ACCESS and a senior member of IEEE.
\end{IEEEbiographynophoto}

\begin{IEEEbiographynophoto}{LONG ZHANG} (zhanglong3211@yeah.net)
is pursuing his Master degree at the School of Communication and Information Engineering, Chongqing University of Posts and Telecommunications, Chongqing, China. His research interests include mobile edge computing and internet of things.
\end{IEEEbiographynophoto}

\begin{IEEEbiographynophoto}{SHAHID MUMTAZ} (smumtaz@av.it.pt)
received the M.Sc. degree from the Blekinge Institute of Technology, Sweden, and the Ph.D. degree from the University of Aveiro, Portugal. He is currently a Senior Research Engineer with the Institute of Telecommunications, Aveiro, where he is involved in EU funded projects. His research interests include MIMO techniques, multi-hop relaying communication, cooperative techniques, cognitive radios, game theory, energy efficient framework for 4G, position information-assisted communication, and joint PHY and MAC layer optimization in LTE standard. He has authored several conferences, journals, and books publications.
\end{IEEEbiographynophoto}

\begin{IEEEbiographynophoto}{ZHENYU ZHOU} (zhenyu\_zhou@ncepu.edu.cn)
received the M.E. and Ph.D. degrees from Waseda University, Tokyo, Japan, in 2008 and 2011, respectively. Since 2013, he has been an Associate Professor with the School of Electrical and Electronic Engineering,
North China Electric Power University, Beijing, China. His research interests include green communications and smart grids. Dr. Zhou served as an Associate Editor for IEEE Access, a Guest Editor for IEEE Communications Magazine, workshop co-Chair for the 2015 IEEE International Symposium on Autonomous Decentralized Systems, and a TPC member for IEEE VTC, IEEE ICC, IEEE Globecom, IEEE PIRMC, IEEE APCC, IEEE Africon, IEEE CCNC, etc.
\end{IEEEbiographynophoto}

\begin{IEEEbiographynophoto}{MUGEN PENG} (pmg@bupt.edu.cn)
received the Ph.D. degree in communication and information systems from the Beijing University of Posts and Telecommunications (BUPT), Beijing, China, in 2005. Afterward, he joined BUPT, where he has been a Full Professor since 2012. His main research areas include wireless communication theory, radio signal processing, cooperative communication, selforganization networking, heterogeneous networking, cloud communication, and Internet of Things. Dr. Peng was a recipient of the 2018 Heinrich Hertz Prize Paper Award, the 2014 IEEE ComSoc AP Outstanding Young Researcher Award, and the Best Paper Award in the JCN 2016, IEEE WCNC 2015, etc.
\end{IEEEbiographynophoto}

\end{document}